*Research Article*

# Hawking Radiation-Quasinormal Modes Correspondence for Large AdS Black Holes

**Dao-Quan Sun, Zi-Liang Wang, Miao He, Xian-Ru Hu, and Jian-Bo Deng**

*Institute of Theoretical Physics, Lanzhou University, Lanzhou 730000, China*

Correspondence should be addressed to Jian-Bo Deng; dengjb@lzu.edu.cn

It is well-known that the nonstrictly thermal character of the Hawking radiation spectrum generates a natural correspondence between Hawking radiation and black hole quasinormal modes. This main issue has been analyzed in the framework of Schwarzschild black holes, Kerr black holes, and nonextremal Reissner-Nordstrom black holes. In this paper, by introducing the effective temperature, we reanalyze the nonstrictly thermal character of large AdS black holes. The results show that the effective mass corresponding to the effective temperature is approximatively the average one in any dimension. And the other effective quantities can also be obtained. Based on the known forms of frequency in quasinormal modes, we reanalyze the asymptotic frequencies of the large AdS black hole in three and five dimensions. Then we get the formulas of the Bekenstein-Hawking entropy and the horizon's area quantization with functions of the quantum "overtone" number $n$.

## 1. Introduction

Ever since Hawking's astounding discovery [1] that black holes radiate thermally, many studies have been carried out [2–11]. One of the works done by Parikh and Wilczek, considering the conservation of energy, shows that an isolated radiation black hole cannot have an emission spectrum that is precisely thermal [2, 3]. Nonstrictly thermal character of Hawking radiation spectrum implies that the spectrum is not strictly continuous and thus generates a natural correspondence between Hawking radiation and black hole quasinormal modes [12–16]. It is very important for the physical interpretation of black holes quasinormal modes. Hod's intriguing idea [17, 18] suggested that black hole quasinormal modes carry important information about black hole's area quantization. Afterwards, Hod's influential conjecture was refined and clarified by Maggiore [19]. In addition, it is also believed that quasinormal modes would be connected with the microstructure of space-time [20]. In [12–16], it was shown that quasinormal modes can be naturally interpreted in terms of quantum levels, where the emission or absorption of a particle is interpreted as a transition between two distinct levels on the discrete energy spectrum. These results are important to realize the unitary quantum gravity theory as black holes are considered as theoretical laboratories for testing models of quantum gravity.

Based on the nonstrictly thermal character, the effective framework of black holes was introduced by [12–16]. For Schwarzschild black holes [12, 14], the physical solutions for the absolute values of the frequencies can be used to analyze important properties and quantities, like the horizon area quantization, the area quanta number, the Bekenstein-Hawking entropy, and the number of microstates [21], that is, quantities which are considered fundamental to realize the underlying unitary quantum gravity theory.

Research showed that asymptotical AdS solutions are stable [22]. The quantum mechanical and thermodynamic properties of black holes in AdS space-time have been considered [22, 23]. Some researches also show that the quasinormal frequencies of AdS black holes could be interpreted in terms of dual conformal field transformation [24–26]. In this paper, we would like to focus on the quasinormal frequencies of large AdS black holes. We apply the effective framework of nonstrictly thermal character to large AdS black holes in different dimensions. We introduce the effective temperature and show that the effective mass is approximatively the average one in any dimension. As examples, we will give three- and five-dimensional cases. Based on the known forms of frequency in quasinormal modes, we reanalyze the asymptotic frequencies of the large AdS black hole in three and five dimensions and calculate the Bekenstein-Hawking entropy, the horizon area quantization, the area quanta number, and the number of microstates.

The paper is organized as follows: in Section 2, we will study the effective temperature for large AdS black holes and the corresponding quasinormal modes. Section 3 is reserved for conclusions and discussions.

## 2. Effective Application of Quasinormal Modes to the Large AdS Black Hole

In this section, we will apply the nonstrictly thermal black hole effective framework of [12–16] to large AdS black holes. Working with $G = c = k_B = \hbar = 1/4\pi\varepsilon_0 = 1$ (Planck units) is adopted in the following. The metric of a $d$-dimensional AdS black hole can be written as

$$ds^2 = -f(r) dt^2 + \frac{dr^2}{f(r)} + r^2 d\Omega_{d-2}^2,$$

$$f(r) = \frac{r^2}{R^2} + 1 - \frac{\omega_{d-1} M}{r^{d-3}},$$

(1)

where $R$ is the AdS radius and $M$ is the mass of the black hole. For a large black hole, the metric function $f(r)$ is simplified to

$$f(r) = \frac{r^2}{R^2} - \frac{\omega_{d-1} M}{r^{d-3}},$$

(2)

where $\omega_{d-1} = 16\pi/(d-2)A_{d-2}$, $A_{d-2} = 2\pi^{(d-1)/2}/\Gamma((d-1)/2)$ is the volume of a unit $(d-2)$-sphere.

In general, the rate of particle emission from the horizon is as follows [2]:

$$\Gamma \sim e^{-\Delta S_{BH}}.$$

(3)

There are two different tunneling methods about calculating the Hawking temperature. Both formulas come from a semiclassical approximation with a scalar field on a curved background to calculate the tunneling amplitude but they differ by a factor of 2 in the resulting temperature. The one method which is called canonically invariant tunneling uses a particular ansatz for the action and then solves the Hamilton-Jacobi equations to find the imaginary part; for details, see [5–11]. In this paper, we will focus on the method proposed by Parikh and Wilczek [2] to calculate the relevant physical quantities. And the form of the probability of the emission for the particle in a $d$-dimensional AdS black hole is given by the following [27]:

$$\text{Im } S = \frac{1}{8} A_{d-2} \left[ r_h^{d-2}(M) - r_h^{d-2}(M-\omega) \right].$$

(4)

For a large AdS black hole, the radius of the horizon is $r_h = [R^2 \omega_{d-1} M]^{1/(d-1)}$. Then, the probability will be

$$\Gamma = \exp\left[-\frac{4\pi}{d-2} (\omega_{d-1})^{-1/(d-1)} \right.$$
$$\left. \cdot R^{(2d-4)/(d-1)} \left( M^{(d-2)/(d-1)} - (M-\omega)^{(d-2)/(d-1)} \right) \right].$$

(5)

By making the second-order Taylor expansion of $\Delta S_{BH}$ about $\omega$,

$$\Gamma = \exp\left[ -4\pi (\omega_{d-1})^{-1/(d-1)} \right.$$
$$\cdot R^{(2d-4)/(d-1)} \left( \frac{1}{d-1} \frac{\omega}{M^{1/(d-1)}} \right.$$
$$\left. \left. + \frac{1}{2(d-1)^2} \frac{\omega^2}{M^{d/(d-1)}} \right) \right].$$

(6)

For a large AdS black hole, the Hawking temperature is $T_H = ((d-1)/4\pi)(r_h/R^2)$. From (6), we transform $\Gamma$ to get

$$\Gamma = \exp\left[ -\frac{\omega}{T_H} \left( 1 + \frac{1}{2(d-1)} \frac{\omega}{M} \right) \right],$$

(7)

and the leading term gives the thermal Boltzmann factors for the emanating radiation. The second term represents corrections from the response of the background geometry to the emission of a quantum.

Then we introduce the effective temperature:

$$T_E(\omega) = \frac{2(d-1)M}{2(d-1)M + \omega} T_H = \frac{2(d-1)\gamma M^{d/(d-1)}}{2(d-1)M + \omega},$$

(8)

where $\gamma = ((d-1)/4\pi)((R^2 \omega_{d-1})^{1/(d-1)}/R^2)$. Then, (7) can be rewritten in Boltzmann-like form:

$$\Gamma \sim \exp\left[-\beta_E(\omega)\omega\right] = \exp\left(-\frac{\omega}{T_E(\omega)}\right),$$

(9)

where $\beta_E(\omega) = 1/T_E(\omega)$ and $\exp[-\beta_E(\omega)\omega]$ is the effective Boltzmann factor. From the Hawking temperature $T_H = ((d-1)/4\pi)(r_h/R^2)$ and (8), we can introduce the effective mass $M_E = [2(d-1)]^{d-1} M^d/[2(d-1)M+\omega]^{d-1}$ and, then, we Taylor expand $M_E$ in $\omega/M$. Thus, for $d$-dimensional large AdS black holes,

$$M_E = M - \frac{\omega}{2}.$$

(10)

As one can see, the effective mass is the average quantity between the states before and after the emission. Here, it should be remarked that the original idea is that, during the particle emission, the effective mass is the average quantity, which has only been proved in Schwarzschild black holes, and can be generalized to cases of the large AdS black hole in any dimension. Thus, in any dimension, we may also introduce the other effective quantities, such as the large AdS black hole the effective horizon $r_E$, effective horizon area $A_E$, the Bekenstein-Hawking large AdS black hole effective entropy $S_E$, and other quantities.

Since the asymptotic forms of quasinormal modes of the 3-dimensional and 5-dimensional large AdS black holes were given by [28], we will reanalyze the asymptotic frequencies of the large AdS black hole in the three and five dimensions by introducing effective temperature.

For the 5-dimensional large AdS black hole, $r_h = (R^2 \varpi_4 M)^{1/4}$, $T_H = (1/\pi)(r_h/R^2) = (R^2 \varpi_4 M)^{1/4}/\pi R^2$. From (8), we can easily get

$$T_E(\omega) = \frac{8M}{8M + \omega} T_H = \frac{8\gamma_5 M^{5/4}}{8M + \omega}, \quad (11)$$

where $\gamma_5 = (R^2 \varpi_4)^{1/4}/\pi R^2$. The asymptotic form of quasinormal modes of the 5-dimensional large AdS black hole is given as follows [28]:

$$\omega_n = 2\pi T_H n (\pm 1 - i). \quad (12)$$

To consider the nonstrictly thermal spectrum of the large AdS black hole, we will correct the asymptotic form of quasinormal modes of the 5-dimensional large AdS black hole. Namely, we substitute the Hawking temperature $T_H$ in (12) by the effective temperature $T_E$. Now, (12) yields

$$\omega_n = \pm 2\pi T_E n - i 2\pi T_E n = m_n + i p_n, \quad (13)$$

where

$$m_n = \pm \frac{16\pi \gamma_5 M^{5/4}}{8M + |\omega_n|} n,$$
$$p_n = -\frac{16\pi \gamma_5 M^{5/4}}{8M + |\omega_n|} n. \quad (14)$$

Now $|\omega_n|$ is the damped harmonic oscillator's proper frequency that is defined as follows [12, 14, 19]:

$$|\omega_n| = \sqrt{m_n^2 + p_n^2}. \quad (15)$$

By using the new expression (13) for the frequencies of quasinormal modes, we get

$$|\omega_n| = \sqrt{m_n^2 + p_n^2}$$
$$= \sqrt{\left(\frac{16\pi \gamma_5 M^{5/4}}{8M + |\omega_n|} n\right)^2 + \left(\frac{16\pi \gamma_5 M^{5/4}}{8M + |\omega_n|} n\right)^2} \quad (16)$$
$$= \frac{16\sqrt{2}\pi \gamma_5 M^{5/4}}{8M + |\omega_n|} n.$$

Then, one can get

$$|\omega_n| = -4M \pm 4M\sqrt{1 + \sqrt{2}\gamma_5 \pi M^{-3/4} n}. \quad (17)$$

Clearly, only $|\omega_n| > 0$ has physical meaning. So, we can get

$$|\omega_n| = 4M\sqrt{1 + \sqrt{2}\gamma_5 \pi M^{-3/4} n} - 4M. \quad (18)$$

Since the emitted energy is much less than the black hole mass, that is, $|\omega_n| < M$; thus we have

$$4M\sqrt{1 + \sqrt{2}\gamma_5 \pi M^{-3/4} n} - 4M < M. \quad (19)$$

Then, we give a maximum value for the quantum "overtone" number $n$:

$$n < n_{\max} = \frac{9R^{3/2} M^{3/4}}{16\sqrt{2}\varpi_4^{1/4}}. \quad (20)$$

The above approach has had some important consequences on the quantum physics of black hole. Then our following discussion will start with the quantization of the black hole horizon area. For the 5-dimensional large AdS black hole the horizon area $A$ is related to the mass through the relation $A = 2\pi^2 (R^2 \varpi_4 M)^{3/4}$, and a variation $\Delta M$ in the mass generates a variation:

$$\Delta A = \frac{3\pi^2 R^{3/2} \varpi_4^{3/4}}{2M^{1/4}} \Delta M. \quad (21)$$

In any case, assuming a transition $n \to n - 1$, (13) gives an emitted energy:

$$\Delta M = |\omega_n - \omega_{n-1}| = f_5(M, n), \quad (22)$$

where we have defined

$$f_5(M, n) = 4M\sqrt{1 + \sqrt{2}\gamma_5 \pi M^{-3/4} n}$$
$$- 4M\sqrt{1 + \sqrt{2}\gamma_5 \pi M^{-3/4} (n-1)}. \quad (23)$$

Therefore,

$$\Delta A = \frac{3\pi^2 R^{3/2} \varpi_4^{3/4}}{2M^{1/4}} f_5(M, n), \quad (24)$$

due to the black hole horizon area associated with Bekenstein-Hawking entropy $S = A/4$; thus

$$\Delta S = \frac{3\pi^2 R^{3/2} \varpi_4^{3/4}}{8M^{1/4}} f_5(M, n). \quad (25)$$

As can be seen, (24) and (25) give the corrected formula of the horizon's area quantization related to the quantum "overtone" number $n$ by introducing the effective temperature. In the approximation of Taylor expansion,

$$f_5(M, n) \approx \frac{2\sqrt{2}\varpi_4^{1/4} M^{1/4}}{R^{3/2}},$$
$$\Delta A \approx 3\sqrt{2}\pi^2 \varpi_4, \quad (26)$$
$$\Delta S \approx \frac{3\sqrt{2}}{4}\pi^2 \varpi_4.$$

We can see that under the condition of approximation the area spectrum is equidistant. Our results of (26) are the same as (29) and (31) of [29] which calculated the area and entropy spectra for 5-dimensional large AdS black holes by the Bohr-Sommerfeld quantization condition to the adiabatic invariant quantity [29, 30].

Now, assuming that the horizon area is quantized with quantum $\Delta A = a$, where $a = (3\pi^2 R^{3/2} \varpi_4^{3/4}/2M^{1/4})f_5(M,n)$, the total horizon area is $A = N\Delta A = Na$, where $N$ is the number of quanta of area. We give

$$N = \frac{A}{\Delta A} = \frac{2\pi^2 (R^2 \varpi_4 M)^{3/4}}{a} = \frac{4M}{3f_5(M,n)}. \quad (27)$$

The above analysis will have important consequences on entropy and microstates. It is usually believed that any candidate for quantum gravity must explain the microscopic origin of the Bekenstein-Hawking entropy.

From (27), we can give

$$S_{BH} = \frac{A}{4} = \frac{3N\pi^2 R^{3/2} \varpi_4^{3/4}}{8M^{1/4}} f_5(M,n), \quad (28)$$

and the formula of the famous Bekenstein-Hawking [1, 31, 32] entropy becomes function of the quantum "overtone" number $n$.

According to [33], the number of microstates for this radiation is simply

$$\Omega_{\text{radiation}}(\omega) = \frac{1}{\Gamma(\omega)}, \quad (29)$$

where $\Gamma(\omega)$ is the radiation probability for the an emission $\omega$. Thus, the number of microstates for a large AdS black hole with $M$ is found to be

$$\Omega_5 = \prod_{i=1}^{n} \frac{1}{\Gamma_5(\omega_i)} = e^{(\pi^2/2)(R^2 \varpi_4 M)^{3/4}}, \quad (30)$$

where $\Gamma_5(\omega_i) = \exp[-(4\pi/3)R^{3/2}(\varpi_4)^{-1/4}(M^{3/4}-(M-\omega_i)^{3/4})]$. From (27) and (30), we can get

$$\Omega_5 = e^{(3N\pi^2 R^{3/2} \varpi_4^{3/4}/8M^{1/4})f_5(M,n)}. \quad (31)$$

We can see that the number of microstates for the large AdS black hole becomes a function of the quantum "overtone" number $n$.

According to discussion of the 5-dimensional AdS large black hole, we will do the same approach to treat the 3-dimensional large AdS black hole.

For the 3-dimensional AdS large black hole $r_h = R(\varpi_2 M)^{1/2}$, $T_H = r_h/2\pi R^2$. From (8) we can easily get

$$T_E = \frac{4\gamma_3 M^{3/2}}{4M + \omega}, \quad (32)$$

where $\gamma_3 = \sqrt{\varpi_2}/2\pi R$. The asymptotic form of quasinormal modes of the 3-dimensional large AdS black hole is given as follows [28]:

$$\omega_n = \pm 4\pi T_H \hat{p} - i4\pi T_H n, \quad (33)$$

where $\hat{p}^2 = p/4\pi RT_H$. To consider the nonstrictly thermal spectrum of the large AdS black hole, we will correct the asymptotic form of quasinormal modes of the 3-dimensional large AdS black hole. Namely, we substitute the Hawking temperature $T_H$ in (33) by the effective temperature $T_E$. Now, (33) yields

$$\omega_n = \pm 4\pi T_E \hat{p}_E - i4\pi T_E n = m_n + ip_n, \quad (34)$$

where

$$m_n = \pm \sqrt{\frac{4\pi p}{R}} \frac{2\gamma_3^{1/2} M^{3/4}}{\sqrt{4M + |\omega_n|}},$$

$$p_n = -\frac{16\pi \gamma_3 M^{3/2}}{4M + |\omega_n|} n. \quad (35)$$

By using the new expression (34) for the frequencies of quasinormal modes, then we get

$$|\omega_n| = \sqrt{m_n^2 + p_n^2}$$
$$= \sqrt{\frac{16\pi p}{R} \frac{\gamma_3 M^{3/2}}{4M + |\omega_n|} + \frac{256\pi^2 \gamma_3^2 M^3}{(4M + |\omega_n|)^2} n^2}, \quad (36)$$

where the solution of (36) in terms of $|\omega_n|$ will be the answer of $|\omega_n|$. Therefore, given a quantum transition between $n$ and $n-1$, we define

$$\Delta M = |\omega_n - \omega_{n-1}| = f_3(M,n). \quad (37)$$

For the 3-dimensional large AdS black hole, the horizon area $A = 2\pi R\sqrt{\varpi_2 M}$ and a variation $\Delta M$ in the mass generates a variation:

$$\Delta A = \pi R \sqrt{\frac{\varpi_2}{M}} \Delta M = \pi R \sqrt{\frac{\varpi_2}{M}} f_3(M,n). \quad (38)$$

Thus,

$$\Delta S = \frac{\pi R}{4} \sqrt{\frac{\varpi_2}{M}} f_3(M,n), \quad (39)$$

$$N = \frac{A}{\Delta A} = \frac{2M}{f_3(M,n)}. \quad (40)$$

From (40), we can give Bekenstein-Hawking entropy:

$$S_{BH} = \frac{A}{4} = \frac{N}{4} \pi R \sqrt{\frac{\varpi_2}{M}} f_3(M,n), \quad (41)$$

which becomes the function of the quantum "overtone" number $n$.

Same as the technique with the 5-dimensional large AdS black hole, we can get the number of microstates:

$$\Omega_3 = \prod_{i=1}^{n} \frac{1}{\Gamma_3(\omega_i)} = e^{(1/2)\pi R \sqrt{\varpi_2 M}}$$
$$= e^{(N/4)\pi R \sqrt{\varpi_2/M} f_3(M,n)}, \quad (42)$$

where $\Gamma_3(\omega_i) = \exp[-4\pi R(\varpi_2)^{-1/2}(M^{1/2} - (M - \omega_i)^{1/2})]$.

## 3. Conclusion

In this paper, by introducing the effective temperature, the analysis on the nonstrictly thermal character of the large AdS black hole is presented. Our results show that the effective mass corresponding to the effective temperature is approximatively the average one in any dimension. And the other effective quantities can also be obtained. Based on the known forms of frequency in quasinormal modes, we reanalyze the asymptotic frequencies of the large AdS black hole in three and five dimensions. Then the formulas of the Bekenstein-Hawking entropy and the horizon's area quantization with functions of the quantum "overtone" number $n$ are got. Moreover, the results we give in the five dimensions have a good consistency with the original one in the approximation of Taylor expansion. We can also see that under the condition of approximation the area spectrum is equidistant.

## Conflicts of Interest

The authors declare that there are no conflicts of interest regarding the publication of this paper.

## Acknowledgments

The authors would like to acknowledge the National Natural Science Foundation of China (Grant no. 11571342) for supporting them during this work.

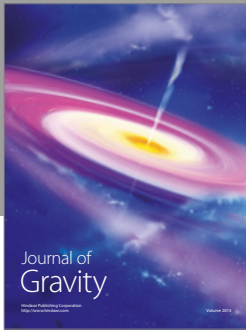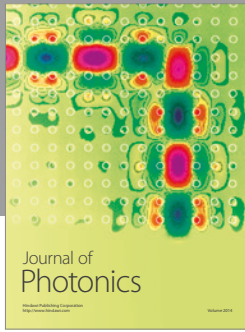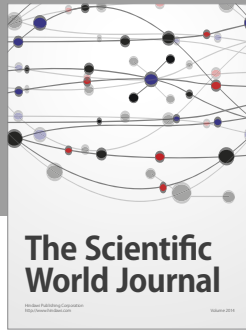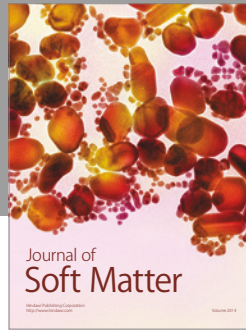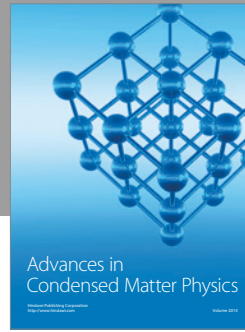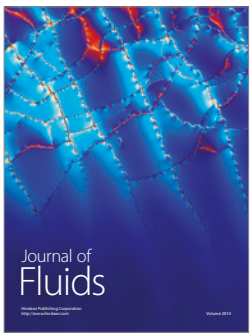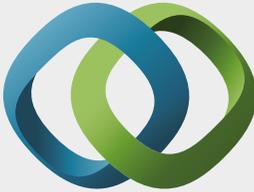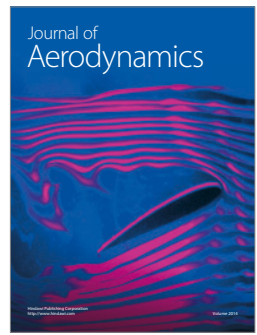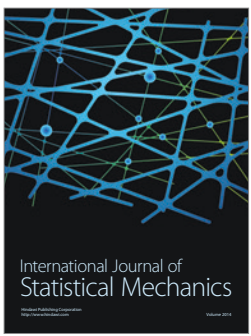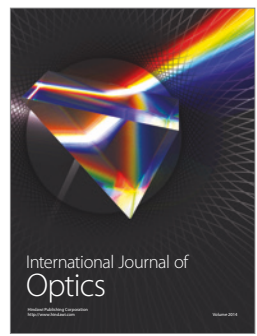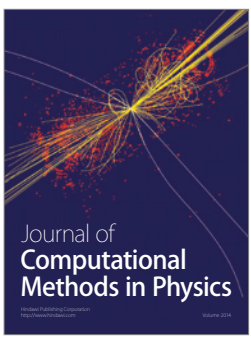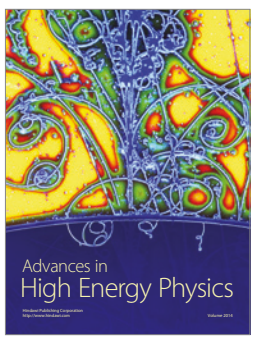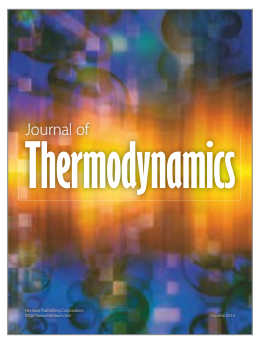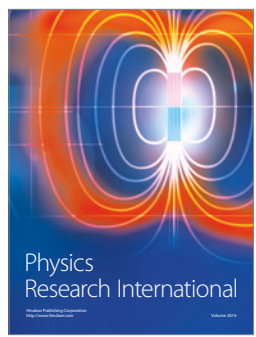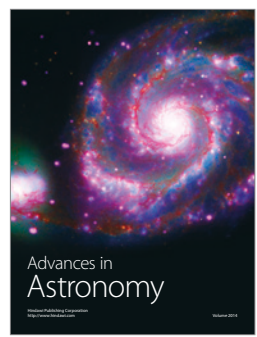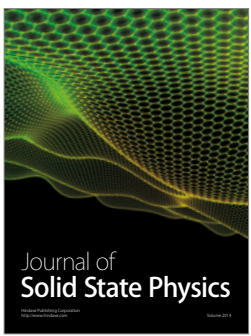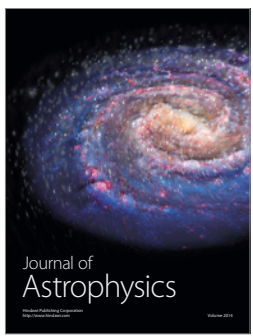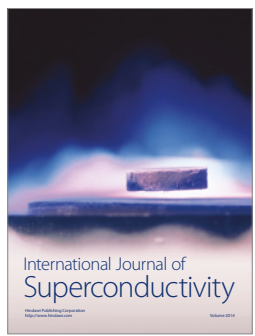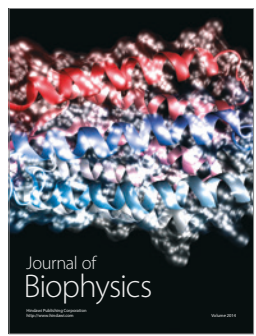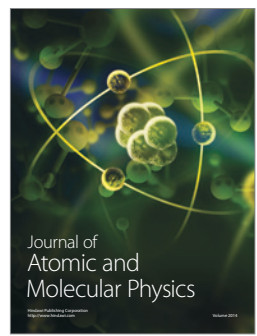